\documentclass{ieeeaccess}
\usepackage{cite}
\usepackage{amsmath,amssymb,amsfonts}
\usepackage{algorithmic}
\usepackage{graphicx}
\usepackage{textcomp}

\def\BibTeX{{\rm B\kern-.05em{\sc i\kern-.025em b}\kern-.08em
    T\kern-.1667em\lower.7ex\hbox{E}\kern-.125emX}}
\begin{document}
\history{This is the authors' accepted manuscript of the article
published in IEEE Access. The final published version is available at
\texttt{https://doi.org/10.1109/ACCESS.2025.3546526}.}

% Use the real DOI here (optional in \doi, but fine to keep)
\doi{10.1109/ACCESS.2025.3546526}

\title{Leveraging the Learning Curve: Reusing Existing Architectural Patterns to Design and Implement MAS}
\author{
\uppercase{Arthur Casals}, \IEEEmembership{Member, IEEE},
and \uppercase{Anarosa A. F. Brandão} %\IEEEmembership{Member, IEEE}.
}
\address{Escola Politécnica da USP, São Paulo, Brazil (e-mail: \{arthur.casals,anarosa.brandao\}@usp.br)}
\tfootnote{This study was financed in part by the Coordenação de Aperfeiçoamento de Pessoal de Nível Superior - Brazil (CAPES) - Finance Code 001}

\markboth
{Casals, Brandão: MAS:DS}
{Casals, Brandão: MAS:DS}

\begin{abstract}
Recent advancements in AI have led to the development of specialized systems related to multi-agent systems (MAS). However, the inherently collaborative nature of agents is often overlooked, and many of these specialized systems are used as components by other AI systems. From a software engineering perspective, this context can benefit from aligning the architectural characteristics of distributed systems with the inherently distributed nature of MAS. We propose that introducing a minimal set of agent-related concepts into the Distributed Systems (DS) domain can improve the engineering of modern MAS by leveraging techniques from DS engineering with established agent theory. In this study, we recapitulated the common origins of MAS and DS by drawing architectural parallels to establish a unified engineering approach. We then defined a minimal set of agent concepts to perform two practical studies on leveraging MAS development. First, we incorporated these concepts into a DS architectural pattern to design a distributed MAS. We then used these concepts in a graduate course to teach MAS engineering to students with no prior knowledge of agent theory. The learning outcomes from both courses included successful MAS implementation using DS tools and techniques. Although more than two-thirds of these students had no practical experience in developing distributed systems, the average final grade in both courses was above 80\%, thus validating our approach. Finally, we discuss how this study supports the development of advanced systems using modern AI techniques consistently with established agent-related research while leveraging established DS techniques and concepts.
\end{abstract}


\titlepgskip=-15pt

\maketitle

\section{Introduction}
\label{sec:introduction}

In 1995, Russell and Norvig~\cite{Russell:1995:AIM:193191} stated that ``AI is the study of agents". The agent concept is also used in Distributed Artificial Intelligence (DAI)~\cite{gasser2014distributed}. Fikes~\cite{fikes1982commitment} first described the basis for autonomous and cooperative problem-solving in 1982, laying the foundation for agents as autonomous entities used for problem solving. 

The main concern of DAI is to find a ``collaborative solution of global problems by distributed entities"~\cite{gasser2014distributed}. In this case, entities can possess different natures and levels of complexity, including \textit{reasoning} - thus characterizing the "intelligent" aspect. The problem-solving process is based on the concept of \textit{information sharing}. \textit{Collaboration} arises from shared information being necessary for entities to solve the problems at hand. Simultaneously, the problem is \textit{global} and common to all the entities involved in the problem-solving process. Finally, the process is considered \textit{distributed} because the entities involved in the problem-solving process can exist in different locations, as determined by logical or geographical separation.

These concepts were also summarized by Demazeau and M\"uller~\cite{muller1990decentralized}, who introduced the concept of \textit{Decentralized Artificial Intelligence} (DzAI). In their article, they state that DzAI ``is concerned with \textit{the activity of an autonomous agent in a multi-agent world}.” The focus of DzAI is to model 
autonomous agents that can solve \textit{their problems} or achieve their goals. 
By being able to possess multiple goals, it is also taken into account that (i) the agent might have to deal with multiple and sometimes contradictory sources of information, and (ii) all of their goals have to be mapped according to the agent's own restrictions in terms of perceiving and acting over the environment. 

Agents can be organized in communities known as \textit{multi-agent systems} (MAS), which are systems composed of multiple agents that interact among themselves in a single environment~\cite{Russell:1995:AIM:193191} to solve complex problems. Regardless of their nature, problems in MAS may possess different degrees of complexity, and they can be divided into subproblems (or goals) individually assigned to each agent in the system~\cite{ferber1999multi}. MAS are often related to or merged with modern DAI concepts due to their origin and characteristics. In 1996, Parunak~\cite{parunak1996applications} stated that structuring a DAI system would require designing both the agent and system architecture, explicitly describing the added value of an MAS in an industrial scenario. The author also stated the benefits of replacing centralized control systems and databases with an MAS whenever a control distribution would be desirable. Modern DAI is categorized as MAS and distributed problem-solving~\cite{gasser2014distributed}. MAS focuses on how agents coordinate their knowledge and activities, whereas distributed problem-solving focuses on decomposing problems and synthesizing solutions. 

Owing to their inherently distributed mechanisms, it is easy to trace a parallel between MAS and \textit{distributed systems}. A distributed system (DS) is ``a collection of autonomous computing elements that appear to its users as a single coherent system"~\cite{steen2017distributed}. Similar to MAS, their components concurrently interact with each other to achieve a common goal. Additionally, their computing elements (\textit{nodes}) can behave independently. 

Among the existing distributed systems currently used on a large scale, the World Wide Web (WWW) is one of the most widely used and developed ones. It was created in 1994 by Tim Berners-Lee as a loosely coupled distributed system for sharing documents~\cite{berners1994world}, focusing on user-generated content and ease of use. A few years later, the author introduced the concept of the Semantic Web~\cite{berners2001semantic}, extending his original work by adding semantics to the existing data format representation. This extension allows information on the World Wide Web to be understandable to humans and software entities. Designing systems for the Semantic Web involves not only creating systems with distributed capabilities but also systems capable of sharing and reusing knowledge, which is also part of the focus of MAS. From this point on, we will refer to the Semantic Web simply as ``the Web" and systems designed for the Web as ``web-based systems.''

Although there are many different architectures for distributed systems~\cite{steen2017distributed}, web-based systems host some of the most complex use cases for the development of distributed systems because they can be designed for a massive number of simultaneous users that are geographically distributed and make use of both local and distributed resources. In addition, DS can be built with web interfaces, so they can either access existing systems or data on the Web or be accessed by other systems using Web protocols. In our context, this is the case for systems such as ChatGPT~\footnote{https://chat.openai.com/} and Midjourney~\footnote{https://www.midjourney.com/}, which can be used by other systems via API access and Discord~\footnote{https://discord.com/} integration. In either case, each system instance is treated as a software agent with different capabilities and responsibilities, similar to a heterogeneous MAS. While there are many plugins and third-party tools that address the use of multiple instances of such systems as a "multi-agent" tool, most of them do not relate to the existing multi-agent research, neither in general nor in its subfields (e.g., coordination, communication, and collaboration).

Designing an MAS from a DS perspective means taking a minimum set of MAS elements and bringing them to the context of a DS. Interestingly, the majority of existing work related to the Web in the MAS domain adopts the opposite perspective: Web technologies (and their related elements) are brought to the context of MAS. We do not intend to argue about the absolute number of web-based distributed system implementation rates when compared with MAS, nor to speculate why or if one is easier to implement than the other. It is a fact, however, that a myriad of tools, libraries, and frameworks are available for the implementation of DS over the Web~\cite{webber2010rest} that encapsulate the specialized knowledge necessary to design distributed systems.

If we emulate the same scenario in the current MAS context, one could argue that the MAS community lacks the abundance of tools or frameworks available for distributed systems. The existing tools and MASs are almost exclusive to the agent community; all extensions are exclusive to other MAS and Agent frameworks, and many of the frameworks have not yet been used to deploy an MAS in production~\cite{mascardifantastic}. In addition, none of the most popular MAS frameworks eliminate the need for specialized knowledge, such as goals and plans, or how the system can be implemented considering different organizations, artifacts, and collaborating agents.

The objective of the present work is to propose an approach that uses both existing MAS research and modern DS engineering techniques to leverage the learning curve currently necessary for developing MAS. To achieve this objective, we need to analyze the common evolution of MAS and DS to establish a minimum set of MAS concepts that can be introduced into the DS domain and effectively used to build new AI systems. 

Once this minimal conceptual set is established, its viability must be tested. This analysis was conducted in two practical studies. First, we used this minimal conceptual set in conjunction with a specific DS architectural pattern to design a distributed MAS. In the second study, we used both the conceptual set and the previous use case to teach graduate students without any previous knowledge about MAS and Agents to design and implement MAS using DS techniques. These courses took place on two different occasions. 

This paper is organized as follows: Section~\ref{sec:Parallel} presents our study on the parallels between DS and MAS, with an emphasis on the architectural and communication aspects. In Section~\ref{sec:Related}, we explore existing work related to our research, focusing on the use of web technologies in conjunction with MAS. Section~\ref{sec:MASDesign} describes our approach in designing MAS in the context of DS, including our approach for determining a minimal conceptual set of topics in agent theory, the incorporation of such concepts into the DS domain, and practical studies performed to explore our approach. This section also includes the learning outcomes of both the courses. We conclude this paper in Section~\ref{sec:Discussion} with our final considerations and future research directions.

%%%%%%%%%%%%%%%%%%%%%%%%%%%%%%%%%%%%%%%%%%%%%%%%%%%%%%%%%%%%%%%%%%%%%%%%%%%%%%%%%%%%%%%%%%
\section{Distributed Systems and MAS}
\label{sec:Parallel}
According to Steen and Tanenbaum~\cite{steen2017distributed}, distributed systems can be divided into \textit{high-performance distributed computing}, \textit{distributed information}, and \textit{pervasive systems}, depending on their purpose. According to the authors, the reasons for designing a DS should meet four specific goals: (i) making resources easily accessible; (ii) making distribution transparent by hiding the fact that resources are distributed across a network; (iii) offering components that can be easily used by, or integrated into other systems; and (iv) being scalable in size, geographic distribution, or number of administrative organizations (following the scalability dimensions proposed by Neuman~\cite{ord1994scale}).

Similarly, Russell and Norvig~\cite{Russell:1995:AIM:193191} stated that MAS is designed to solve problems that are beyond the capabilities of individual agents or monolithic systems. As mentioned before, multi-agent systems are inherently distributed, solving problems through the cooperation and interoperation of multiple agents. Individual agents can be used not only for distributed task allocation but also for wrapping and providing access to legacy systems. Agents can also use distributed resources to retrieve and coordinate information from multiple sources. In addition, agents are highly reusable and composable~\cite{bergenti2002discussion}.
Similar to distributed systems, MAS can be built for high-performance computing tasks~\cite{leitao2013parallelising,rousset2016survey}, disseminating information across multiple nodes~\cite{chen2000distributed}, and for ubiquitous blending with the environment~\cite{su2011jade}. 

We trace a parallel between DS and MAS in this paper because while they share multiple elements in their history, both are seen and used differently by the software engineering community. In particular, Web-based distributed systems are widely used and adopted by many software engineers and organizations across the globe, whereas MAS is not yet mainstream. While we do not have the tools (or data) to affirm that this is because of one reason or another, it is also a fact that the existing resources related to Web-based DS are \textit{distributed systems implemented from a Web perspective}. This means that elements related to the DS research field were introduced and implemented in the context of the Web. Because the Web itself evolved as a distributed system, it is difficult to define the line that separates bringing one context into another. However, it is relatively easy to perceive that most distributed systems theory is shielded from the average web systems engineer. From this perspective, taking MAS elements and bringing them to the context of DS engineering could potentially leverage the development of MAS.

In the following paragraphs, we analyze the main aspects of designing a DS and establish a parallel with the MAS design aspects. These aspects revolve around two specific topics: (i) architecture and (ii) communication. One characteristic common to both DS and MAS is that all the other engineering aspects - from coordination to handling failures - depend on the system's architecture and communication. It can be argued that "available resources" (hardware, auxiliary systems, communication bandwidth, etc.) are also a primary topic when designing MAS and DS. However, because resource analysis is a fundamental part of designing the system architecture in the first place, we do not need to address it. This analysis aims to create a baseline for existing work in both domains, which will later be used to establish the minimal MAS conceptual set mentioned previously. The DS aspects analyzed were based on the organization proposed by Steen and Tanenbaum~\cite{steen2017distributed}.
%%%%%%%%%%%%%%%%%%%%%%%%%%%%%%%%%%
\subsection{Architecture}
Software architectures represent how software components are logically organized and how they should interact with each other~\cite{bass2003software}. They use architectural styles to organize system components and make their inherent complexities manageable and understandable. An architectural style ``is formulated in terms of components, the way in which components are connected to each other, the data exchanged between components, and finally how these elements are jointly configured into a system"~\cite{steen2017distributed}. The components and their relationships follow the definition provided by Selic~\cite{selic2004,selic2006}, being modular and providing replaceable interfaces within the environment. In general, connectors mediate everything related to communication between components, including information and control flow. 

\subsubsection{Architecture in DS}
According to Steen and Tanenbaum~\cite{steen2017distributed}, the most important architectural styles used when designing distributed systems are (i) layered architectures, (ii) object-based architectures, (iii) resource-centered architectures, and (iv) event-based architectures. These styles can be combined within a system, with each part of the system having a different architectural style. Regardless of the architectural style used, DS can be organized in different ways according to the coordination of its components. \textit{Centralized} DSs rely on a central \textit{coordinator}, which is usually responsible for implementing all the data processing. This coordinator is accessed by a \textit{client} who requests and uses the information provided by the coordinator to perform its functions. 

In contrast, a \textit{decentralized} DS does not possess a unique component that is solely responsible for data processing. Instead, each component of the system possesses its own data processing responsibilities, and they can communicate with each other without the need for third-party intervention. Centralized and decentralized architectures can be combined into hybrid architectures. 

We will not discuss each style in detail (although their names provide a general idea of what is involved in each one). However, we decided to focus our analysis on decentralized DS because of their nature. From an MAS perspective, building a distributed heterogeneous MAS can involve almost all sub-research fields within agent theory, which is ideal when proposing a minimal conceptual set related to a research area. In addition, as mentioned before, there are many different architectures for distributed systems, and web-based systems host some of the most complex use cases for DS development. If we want to consider systems that are partially exposed to the Web (either by deployment or by accessing another web-based system), we need to understand the architectural styles used when building web-based systems.

\subsubsection{Web architectural styles}
In DS, layered architectures are based on organizing software components in layers so that each layer can access the adjacent layers. Top-down layered architectures (i.e., vertically organized) are commonly used in network communication. In this organization, the upper layers can access those immediately below using \textit{calls}. Layered architectures can be organized in different ways~\cite{steen2017distributed}. 

Object-based architectures follow the same principle as layered architectures by dividing software components into objects that can access each other. Each object encapsulates the data and the operations that can be performed on that data. Similarly, services can also be encapsulated and communicate with each other, which is the basis for service-oriented architectures (SOA)~\cite{perrey2003service}. A system designed following SOA principles is composed of multiple services that do not necessarily belong to the same administrative organization or reside on the same server (being, therefore, scalable). Owing to organizational requirements, services can be grouped and divided into layers. These architectural styles are used on the web as \textit{web services}~\cite{newcomer2005understanding}. 

Resource-based architectures are based on the idea that a DS is a collection of resources individually managed by components. In the context of the Web, such architectures are seen as an evolution from pure service-based ones. Resources are considered to be ``the fundamental building blocks of Web-based systems"~\cite{webber2010rest}. Resource-based architectures still depend on services to access and manipulate resources within a system.

Following the previous architectural style, event-based architectures~\cite{garlan1993introduction} are based on the concept of decoupling processes. Steen and Tanenbaum~\cite{steen2017distributed} use the term "publish-subscribe" to refer to the same idea, since the publisher-subscriber messaging pattern~\cite{birman1987exploiting} is commonly used for communication (although event streaming~\cite{carney2002monitoring} is also another communication pattern used in event-based applications).  Coordination and integration between services are achieved through the use of events representing changes in the state of data (such as resources), instead of requiring that all services involved in a process know the full specification of the data involved. Publish-subscribe architectures~\cite{eugster2003many} are examples of this architectural style: a given service (publisher) sends a message notifying other services (subscribers) that something has changed in relation to the state of the data. Other services, on the other hand, react and take action upon this warning, without the need to transfer data between services.

Any of these styles can be used to build and organize distributed systems on the Web. However, the final system architecture also depends on the components it possesses and how they interact with each other, as it reflects how the system is organized~\cite{bass2003software}. Web distributed systems (Web DS), or any other DS, can assume a \textit{centralized} organization (i.e., client-server systems), a decentralized organization (i.e., peer-to-peer systems), or a hybrid between the two (i.e., edge-server systems)~\cite{steen2017distributed}. 

Detailing each type of system architecture used by a DS is not relevant to the present study. However, it is important to highlight that different architectural styles can be used as the basis for systems with different organizations (architectures). Different components (using different technologies) can be organized differently, resulting in different system architectures. We will explore this idea later in the text.

\subsubsection{Middleware}
Another important topic related to DS is \textit{middleware}. In~\cite{krakowiak2007middleware}, Krakowiak defines middleware as the software layer that lies between the operating system and the applications on each site of the system. The main objective of middleware is to mask the heterogeneity and distribution of the underlying hardware and operating systems by providing common programming abstractions, thus hiding low-level programming details. He also presented four distinct design patterns for distributed object middleware (i.e., middleware used in distributed architectural styles other than layered ones):

\begin{enumerate}
    \item [$\bullet$] Proxy: intermediates requests between distributed objects (clients/services) such that the objects do not need to know their mutual locations.
    \item [$\bullet$] Factory: used to dynamically create families of related objects.
    \item [$\bullet$] Adapter: used to provide a different interface for the functions of a specific object (service) in order to comply with the interface expected by another object.
    \item [$\bullet$] Interceptor: used to enhance an existing service with new capabilities or to provide it with different means.
\end{enumerate}

Steen and Tanenbaum~\cite{steen2017distributed} used the adapter and interceptor patterns when describing middleware organization in distributed systems, focusing on how they modify an existing service. While an adapter (or \textit{wrapper)} transforms the service's interface, the interceptor transforms its functionality. These patterns can also be used to provide access to existing web systems or to establish an interface for web access.

Middleware support for MAS is usually limited, providing support for communication or a broker infrastructure to be used by agents~\cite{weyns2010architecture}. According to Weyns, common middleware services such as security, persistency, and transactions are often minimally considered in MAS development. This is because domain-specific middleware for MAS usually considers agent communication as the basis for agent coordination. Other concerns are delegated to domain-independent middleware services. This tendency has been verified in popular frameworks such as JADE~\cite{bellifemine2007developing}, Jack~\cite{wood2000overview}, and JaCaMo~\cite{boissier2013multi}.

\subsubsection{MAS architectures}

According to~\cite{weyns2010architecture}, multi-agent systems have the following characteristics: (i) each agent has incomplete information or capabilities to solve the problem at hand (implying a limited viewpoint), (ii) there
is no system global control, (iii) data is distributed, and (iv) computation is asynchronous. 

Regarding implementation, MAS architectures can follow the same styles as those used in DS. However, at the same time, they also satisfy properties related to agents. Agent theory divides MAS architectures into \textit{deliberative architectures} and \textit{reactive architectures}~\cite{wooldridge1994agent} (which can also be combined into hybrid architectures). Deliberative MAS focuses on (i) translating the environment into a useful description and (ii) \textit{reasoning} upon this information and acting appropriately. Reactive MAS, on the other hand, does not use complex symbolic reasoning, and the agents simply react to the environmental information according to a predetermined program.

The parallel between the DS and MAS architectures involves both implementation and agent theory. From the implementation perspective, MAS can use the same architectural styles as DS, as long as agent theory is satisfied. From the agent theory perspective, an MAS possesses distinct elements (agents, environment, and environment resources) and related processes (planning, perceptions, and actions). These elements can also vary according to the agent paradigm adopted; agents based on the belief-desire-intention (BDI) architecture~\cite{rao1995bdi} require belief revision processes. Without diving into all existing agent paradigms, it is sufficient to say that as long as the elements required by the agent theory are present, an MAS can use the same architectural styles used by a DS.

%%%%%%%%%%%%%%%%%%%%%%%%%%%%%%%%%%
\subsection{Communication}
In distributed systems, the concept of processes is derived from the field of operating systems, where it generally represents ``a program in execution". Considering that distributed systems are composed of processes, ``communication" usually refers to interprocess communication. In MAS, communication occurs between agents either via an agent communication language (ACL) or through the environment.

\subsubsection{Communication in DS}

All communication models in DS are based on message passing: the differences between each model are based on the message type (what is passed between processes) and how it is done. DS communication models are divided into (i) remote procedure calls (RPC), (ii) message-oriented communication, and (iii) multicast communication.

Remote procedure calls allow processes to invoke procedures from other processes. Despite its name, a message is still passed between processes, containing the name of the procedure invoked and its eventual parameters. This communication is usually performed synchronously, in which case the calling process is halted until a response is received. Message-oriented communication, on the other hand, relies solely on messages passed between processes. Each message can contain a full description of what is required from the other process, and it can also be persisted or treated as an independent entity. This can be illustrated by message-queuing systems that support asynchronous communication between processes by handling and persisting messages. Because message-oriented can be either synchronous or asynchronous, message-queuing systems allow processes to communicate with each other even if they are not available simultaneously. Finally, multicast communication is based on the idea of diffusing information from one process to many other processes. Multicasting is usually asynchronous and can also involve message queuing.

\subsubsection{Communication in MAS}
In MAS, communication between agents is also achieved through the use of messages. More specifically, an agent communication language is used to establish a structure for the messages exchanged between agents, both in type and meaning. Agents are also capable of engaging in conversations~\cite{labrou2001standardizing}. A conversation can be seen as a pre-arranged protocol or message exchange pattern oriented towards a specific task or objective.

There are different ACLs in place, but the two most consistently used~\cite{Li2013,kamdar2018state} are the FIPA-ACL~\cite{o1998fipa} and KQML~\cite{Finin:1994:KAC:191246.191322}. Both languages are based on the \textit{speech acts theory}~\cite{searle1969speech} and are composed of different \textit{performatives}.~\footnote{https://stanford.library.sydney.edu.au/entries/speech-acts/}. They can be understood as sentences that describe an environment and influence it simultaneously. As such, messages exchanged between agents can represent actions or communicative actions. Despite being relatively old, these standards are still used~\cite{ruaileanu2018design,blos2018framework}. 

%%%%%%%%%%%%%%%%%%%%%%%%%%%%%%%%%%
\section{Related work}
\label{sec:Related}
Agents have been used in conjunction with DS and Web technologies for a few decades~\cite{jennings1998roadmap,lieberman1995letizia,etzioni1996moving,ardissono1999agent,greenwood2004engineering,dickinson2005agents,ricci2010platform}. In particular, from the Web perspective (both accessing and providing access using web-based technologies), agents have been commonly used as assistants for web-based services~\cite{ardissono1999agent,lieberman1995letizia}. In parallel, considering the use of Web technologies in the MAS domain, agents can also be used \textit{as} web services~\cite{hendler2001agents,huhns2002agents}. Dickinson and Wooldridge~\cite{dickinson2005agents,bai2006multi,hirsch2011education} also proposed the use of cognitive agents as web services. There is also existing work on MAS~\cite{muldoon2007agent,thiele2009mams,tapia2009fusion,casals2018exposing} and agent architectures~\cite{tolk2009agents,paletta2014scouting} using web services. 

The possible beneficial implications of establishing a parallel between MAS and DS have already been discussed previously~\cite{chopra2021multiagent,albrecht2020autonomous,chopra2022interaction,lange1998mobile,ahmad2002multi} from different perspectives. In one of these discussions~\cite{chopra2021multiagent}, the author discusses the problem of the lack of direct impact of MAS on system development practice, pointing towards a possible study of the parallels between DS and MAS. This topic was also a point of discussion between many researchers from the community before~\cite{Mascardi:2019:EMS:3310013.3322175}, which resulted in a study of the current status of existing tools for aiding the development of MAS~\cite{mascardifantastic}.

From a Software Engineering perspective, existing model-driven development (MDD) techniques for developing agent models~\cite{wooldridge2000gaia,bresciani2004tropos,padgham2002prometheus} can also be used to model web-based agents~\cite{zinnikus2008model,kardas2009model,hahn2010enhancing}. In 2007, a survey on existing agent methodologies based on the agent-oriented paradigm was published~\cite{cabri2007service} with the aim of evaluating the extent to which existing agent-oriented methodologies were also oriented toward the development of services. From an Agent-oriented Software Engineering (AOSE) perspective, the study demonstrated that all evaluated AOSE methodologies could be used to model service-oriented agents. At this point, the focus was on modeling agents as services or generally using concepts around services to facilitate the interaction between agents. 

In 2010, however, Ricci \textit{et al.} introduced a platform called CArtAgO-WS~\cite{ricci2010platform}, intended to allow the development of service-oriented applications (SOA) populated by agents. This was a more embracing approach from the MAS perspective, since it took into account both the concepts of artifacts - "objects" or services that the agents could use - and the use of heterogeneous agent architectures (including BDI). This idea was further explored and used with other agent frameworks, such as JaCaMo~\footnote{http://jacamo.sourceforge.net/}. More recently, Ciortea \textit{et al.}~\cite{Ciortea:2017:GAR:3091125.3091342} proposed another step in the same direction, suggesting that the WWW, in its current state, could be suitable as middleware for Internet-scale multi-agent systems. This is achieved through the use of a resource-oriented layer that allows the application environment to be decoupled from its deployment context. To demonstrate this proposal, an agent environment was developed using JaCaMo. This environment was specifically aimed at the Internet of Things (IoT), emphasizing the separation between the application environment and the deployment context.

If we focus on work related to exposing agents as web services, most existing frameworks and implementations aim to expose agents' capabilities as web services. This is done either by exposing each of its capabilities as a service~\cite{greenwood2004engineering} or by exposing the agent as a service that can be accessed through the web and perform its functions~\cite{lieberman1995letizia}. This is what happens when agents are exposed as services using the JADE framework~\footnote{https://jade.tilab.com}, for instance~\cite{casals2018exposing}: while the agents can be accessed through web-compliant addresses, they are still implemented in the context of a MAS, using an MAS framework (JADE), being compliant with an ACL (FIPA-ACL), and so on. 

In this case, web services are used to make the MAS more flexible in terms of accessibility and usage; however, system organization, modeling, and implementation still occur in the MAS context. This is, of course, by design: Dickinson and Wooldridge argued that agents and web services are distinct, since 'agents provide a distinctive additional capability in mediating user goals to determine service invocations'~\cite{dickinson2005agents}. However, maintaining the MAS context while adopting web technologies tends to be the norm in other areas related to MAS. In a recent study, for example, Ricci et al.~\cite{ricci2019engineering} explored the problem of engineering distributed environments and organizations for MAS in a scalable manner. The problem in question is explicitly mapped from the agent perspective onto the web: the MAS is represented as a web resource, and each of its related structures (artifacts, agents, etc.) becomes a sub-resource. In this scenario, agents work with artifacts by accessing their related endpoints through web requests (using REST).

It is interesting to note, however, that while different research has been performed on web-based MAS, most of this work perceives the Internet environment as a means of communication. Even larger proposals such as Agentcities~\cite{willmott2001agentcities} and distributed platforms for data processing~\cite{o2015distributed} rely on web protocols and agent communication standards to establish and coordinate communication processes. In contrast, agents, although exposed as services and sharing resources, are reasoning blocks that take advantage of such a structure. In addition, lower-level communication mechanisms are still based on procedure calls, not taking advantage of newer mechanisms (such as asynchronous message passing). Very recent work in this field aims to address such issues~\cite{Ciortea:2017:GAR:3091125.3091342,ciortea2017beyond,casals2019resource,ricci2019engineering}. 

In the current state of the Web, using it as middleware for MAS is appealing. At the same time, modeling MAS while taking full advantage of the web environment (not only as a communication-enabled environment) is partially subject to the constant evolution of the Internet itself. From a higher-level perspective, it is possible to model a purely web-based MAS using existing AOSE methodologies (such as GAIA~\cite{zambonelli2003developing} and O-MASE~\cite{deloach2004mase,deloach2014mase}) while abstracting the Internet environment. From a lower-level perspective, however, different communication protocols, Web-specific components, frameworks, and technologies can influence the way a distributed MAS can be designed using Web-related technologies.

\section{Distributed Systems and MAS Design} 
\label{sec:MASDesign}
To illustrate how MAS can be implemented from a DS perspective, we adopt a specific architectural pattern called Entity Component System (ECS). One of the main reasons for choosing this pattern is the use of a \textit{data-oriented design} (DOD)~\cite{fabian2020data}. The principles behind DOD have been around for a while, but they received this name in 2009~\footnote{https://gamesfromwithin.com/data-oriented-design}. DOD is a program optimization approach that focuses on the efficient usage of the CPU cache for manipulating data. As such, it emphasizes the data layout, separating it from the problem domain (data is separated from logic). Object-oriented design (OOD)~\cite{meyer1997object}, in contrast, focuses on encapsulating data and attributes within objects. Consequently, the data are inherently contextualized (making it part of the problem domain).

Different studies have been conducted on the advantages of adopting DOD instead of OOD in high-performance systems~\cite{mironov2021comparison,fedoseev2020case}. One of the common conclusions of these studies is that using DOD makes it easier to scale systems that process large quantities of data. This is particularly interesting for MAS development because, in a multi-agent system, there can be, hypothetically, a myriad of agents simultaneously performing actions based on the same data (environment). DOD is frequently used in game development, and it is also interesting from the MAS perspective. In both MASs and games, the system must be designed to handle multiple autonomous entities (agents and game characters). In game development, non-playing characters (NPCs) frequently interact with each other and with the environment. Their actions can depend on the environment's state and affect it as a result (most of the time in real-time). Many other aspects of an agent's field can also be present in game development (e.g., behavioral models, coordination, cooperation, and communication). As a result, when looking for a DS architectural pattern to test our hypothesis, it makes sense to examine the game development field.

\subsection{The Entity-Component-System pattern}
Within this context, one of the most used DS architectural patterns in complex, massive multiplayer online games (MMOGs) is the Entity Component System (ECS)~\cite{nystrom2014game,gregory2017game}. This pattern, based on the concept of Component-Based Systems~\cite{crnkovic2002building}, uses DOD. It appeared and evolved within the MMOG community~\footnote{http://t-machine.org/index.php/2007/09/03/entity-systems-are-the-future-of-mmog-development-part-1/}~\footnote{https://unity.com/dots} and was quickly adopted by modern real-time interactive systems (RIS) frameworks~\cite{wiebusch2015decoupling} and commercial game engines~\cite{baron2019hands}. As a DOD-focused pattern, it emphasizes composition over inheritance by decoupling data from the logic. It is particularly designed for scenarios in which many different, independent entities interact dynamically with the environment and each other. 

The ECS pattern is based on three core abstractions:
\begin{itemize}
    \item[$\bullet$] \textit{Entities}: A general-purpose object or actor within the system. They are usually unique identifiers used by the system and thus contain no logic or data. Their purpose is to serve as a reference point for the \textit{components} (see below). In a game (or MAS), an entity can represent an agent or specific element in the environment.
    \item[$\bullet$] \textit{Components}: Core data structures used by ECSs. Each component stores a single type of data and, as its name implies, they are used solely as data containers, holding no behavior. A single Entity may have multiple attached Components to define its characteristics. In an ECS implementation, Components are stored in arrays or contiguous blocks of memory to promote cache usage efficiency. Because Components are responsible only for storing data, any modifications to the data structures or their components do not affect the overall logic of the program.
    \item[$\bullet$] \textit{Systems}: Logic or behavior used within the program. Systems are designed to operate on Entities with specific Component compositions. Within a game, for example, a System for processing "movement" can be used to update all Entities that possess the combination of components related to "position" and "velocity." 
\end{itemize}

\begin{figure*}
\centerline{\includegraphics[width=\textwidth]{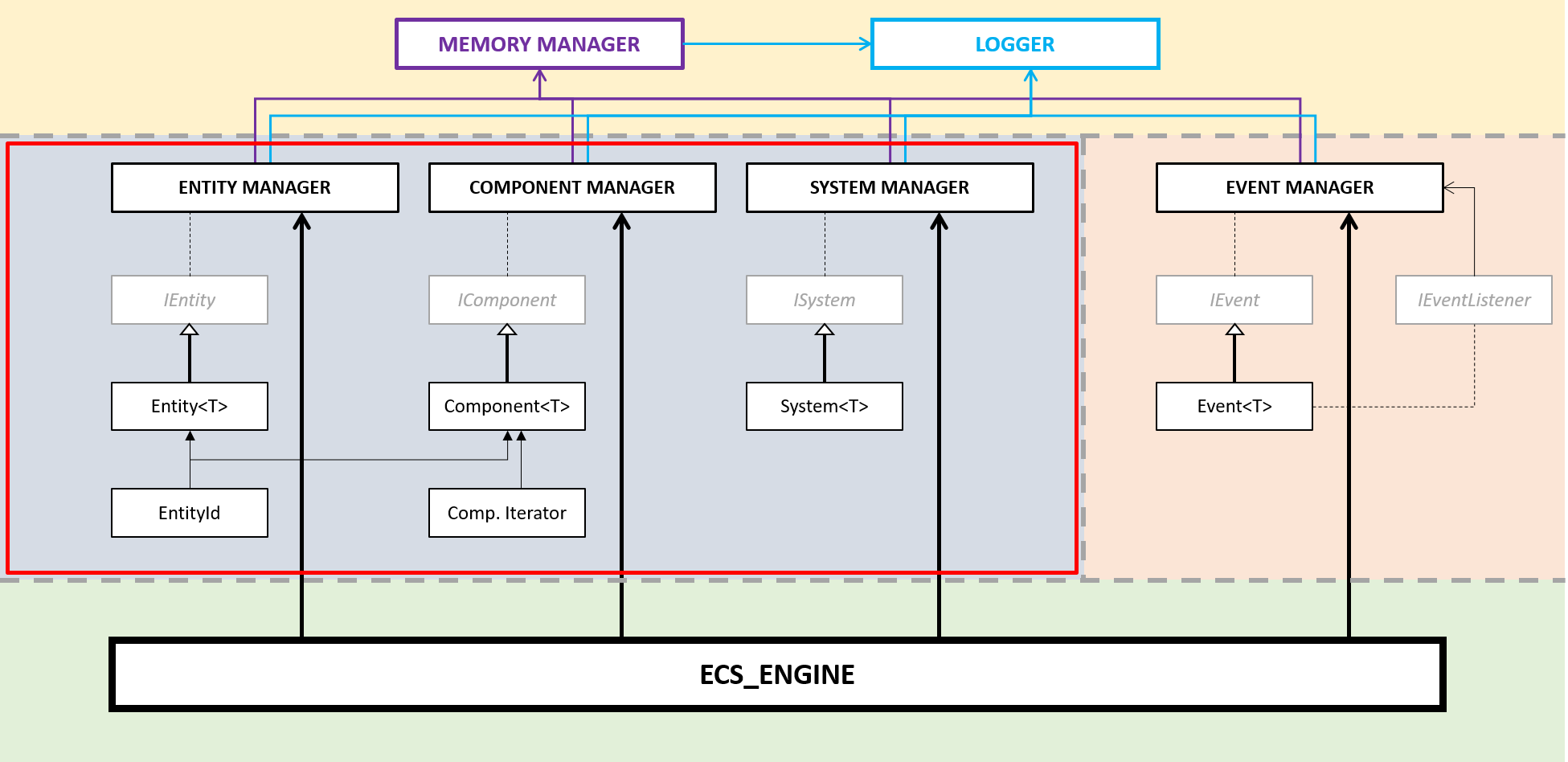}}
\caption{A typical ECS architecture}
\label{fig:ecs}
\end{figure*}

This structure is heavily aligned with the DOD principles, emphasizing data organization for optimal processing efficiency. ECS implementations also use the concepts of (i) ``World" (the environment in which the game happens) and (ii) Events to aid the logic behind systems: all entities are susceptible to different events triggered within the world. A Component is extensible and modifies each Entity by providing specific data in the form of properties. Considering a game environment, a given Entity may possess a Component called "Position," with properties (X, Y, and Z coordinates) that determine where in the world the Entity is located. The organization of a typical ECS is illustrated in Fig.~\ref{fig:ecs}.

\subsection{Agents: a minimal conceptual set}

We also need to define a minimal set of agent-related concepts to incorporate into the adopted architectural pattern. For this task, we decided to use an existing university course as a reference: the introductory MAS course offered by University of São Paulo (USP), Brazil~\footnote{https://edisciplinas.usp.br/course/view.php?id=61282}. 

Our reasoning was based on two main points. The first refers to the course's public. It was designed for graduate students with no prior knowledge of agent theory. Since our goal involves transferring MAS concepts into a different domain, we benefited from the previous years of curriculum fine-tuning and used the course's structure as a basis. The second point involves the course structure; although mostly conceptual, the course involves a reasonable amount of practical activity, including working with existing MAS frameworks. Minimal knowledge of computer programming is required to do so, which also benefits us, considering that the first ones to benefit from our initiative are DS system engineers and architects.

Considering the points above, we also needed to determine whether the course chosen as a reference was aligned with other university courses on the same topic. This analysis is necessary because our main objective involves leveraging the learning curve necessary to develop MAS. Assuming that agent theory is primarily taught in graduation courses within universities, comparing the structure of the course chosen as a reference against a larger pool of agent-related courses is an adequate verification of its suitability for our intended purposes.

We divide our analysis into two parts. In the first part, we analyzed the syllabi of twenty-six graduation courses related to MAS in two different geographical contexts: local (Brazil) and global. This division was intended to show any potentially relevant differences in terms of the curriculum structure. 

\begin{figure}[htbp]
\centering
\includegraphics[width=0.5\textwidth]{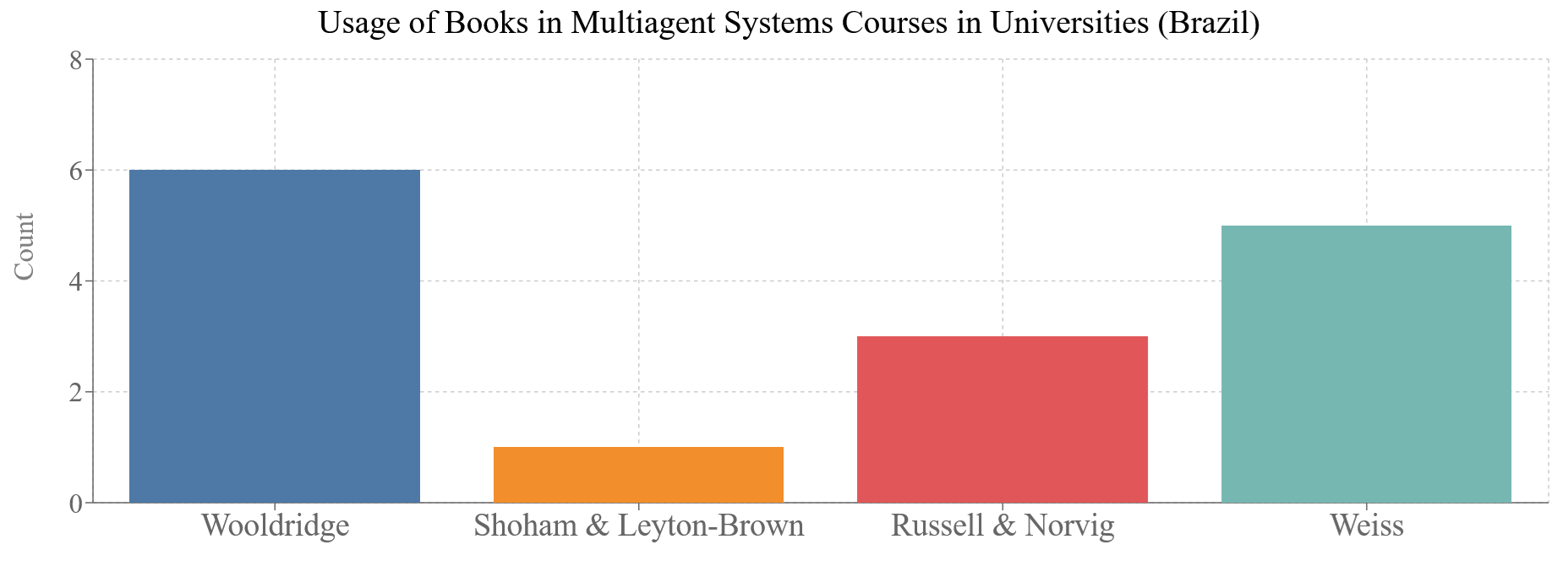}
\caption{Usage of Books in Multi-Agent System Courses in Brazilian Universities (Source: authors)}
\label{fig:brazil_comparison}
\end{figure}

By analyzing these courses, we observed that there were four prevalent books in all syllabi: Wooldridge's~\cite{wooldridge1994agent}, Shoham and Leyton-Brown's~\cite{shoham2009multiagent}, Russell and Norvig's~\cite{Russell:1995:AIM:193191}, and Weiss's~\cite{weiss1999multiagent}. From a global perspective, we observed that approximately 81\% of them referenced Wooldridge's book, whereas others were much less referenced. Russell and Norvig's book, a common reference for the basic definitions of agents, was referenced in only 54\% of all courses. The numbers maintained a similar proportion from a local perspective, except for an increase in the use of Weiss's book by universities in Brazil. This comparison also included the course chosen as a reference. More details of this analysis can be found in Appendix~\ref{app:books}. Table~\ref{tab:comparative_region} shows the comparative usage of MAS courses offered by universities worldwide, and Figures~\ref{fig:brazil_comparison} and~\ref{fig:global_comparison} show the distribution of the four main books cited above among both Brazilian and worldwide universities.

The second part of our analysis relates to the structures of the courses. Assuming that each course is structured based on its syllabus, we analyzed the contents of the three books mentioned above to extract a common set of agent-related concepts. We also cross-referenced these concepts with our previous DS and MAS architecture analyses and the latest Computer Science Curricula prepared by the ACM/IEEE-CS/AAAI Joint Task Force~\cite{10.1145/3664191}. This analysis resulted in the following list of concepts:

\begin{itemize}
    \item[$\bullet$] Basic Definitions: agents, different types of agents (reactive, cognitive, hybrid), MAS.
    \item[$\bullet$] Agent architectures: cognitive agents, reactive agents, BDI.
    \item[$\bullet$] Agent-based communication: message passing and ACLs.
    \item[$\bullet$] Distributed problem-solving: problem decomposition, solution synthesis.
    \item[$\bullet$] Cooperation between agents: cooperative strategies, coordination mechanisms.
    \item[$\bullet$] Task planning: planning algorithms, dynamic task allocation.
    \item[$\bullet$] Coordination between agents: coordination protocols, synchronization techniques.
    \item[$\bullet$] Negotiation between agents: negotiation strategies, conflict resolution.
    \item[$\bullet$] Organizations, rules, and norms: organizational structures, normative systems.
    \item[$\bullet$] Fault Tolerance: fault detection and recovery, redundancy strategies.
    \item[$\bullet$] Learning and Adaptation: machine learning in MAS, reinforcement learning for agent adaptation.
    \item[$\bullet$] Scalability and Performance: scalability techniques, performance optimization.
    \item[$\bullet$] Security and Privacy: secure agent communication, privacy-preserving mechanisms.
    \item[$\bullet$] Human-agent Interaction: interfaces for human-agent interaction, user experience, and usability.
    \item[$\bullet$] Agent-Based Simulation: simulation environments, applications of agent-based modeling.
    \item[$\bullet$] Interoperability: standards and protocols for interoperability, integration with other systems.
    \item[$\bullet$] Autonomous Decision-Making: decision-making algorithms, autonomy levels.
    \item[$\bullet$] Ethics and Trust: ethical considerations in MAS, building trust in agent systems.
    \item[$\bullet$] Resource Management: resource allocation strategies, efficient resource utilization.
    \item[$\bullet$] Advanced AI Techniques in MAS: integrating generative AI and applying deep learning in MAS.
\end{itemize}

In the list above, "BDI" refers to Belief-Desire-Intention agent architectures~\cite{rao1995bdi}, and "ACLs" refers to Agent Communication Languages~\cite{wooldridge1994agent}. This preliminary list covers our observations so far, with a list of MAS-related topics that we judged essential to anyone designing a MAS. 

It is also interesting to observe that the list above shows high synergy with distributed systems. In DS, architectural patterns provide a structured way to organize components, manage communication, and ensure aspects such as scalability and fault tolerance. The MAS architecture is built on the same principles, extending them with intelligent agents capable of making autonomous decisions, learning, and coordinating among themselves.

From the same perspective, not all of these topics are necessary to design every MAS. Depending on the system's requirements, some might not be used at all. Advanced AI techniques will not necessarily be present in every MAS, and purely architectural topics (such as fault tolerance) will not necessarily be addressed, considering the number of frameworks and service providers that encapsulate such concerns. We also need to remember that we want to use this conceptual set in a practical study that involves creating a course on engineering MAS. Therefore, the information taught in such a course must be adequate. Ultimately, we aim for a \textit{minimal conceptual set}.

With the considerations above, the resulting minimal conceptual set is shown below:

\begin{itemize}
    \item[$\bullet$] Basic definitions: agents, different types of agents, MAS.    
    \item[$\bullet$] Agent architectures: cognitive agents, reactive agents, BDI.
    \item[$\bullet$] Agent-based communication: message passing and ACLs.    
    \item[$\bullet$] Cooperation between agents.
    \item[$\bullet$] Task planning.
    \item[$\bullet$] Coordination between agents.    
\end{itemize}

We excluded some engineering aspects from this list, such as "Security" and "Fault Tolerance." Although there are specific sub-fields for this kind of research within the agent research field, we believe that these topics should be considered as part of the architectural requirements (similar to resources). Therefore, they do not belong to this list.

\subsection{Bringing MAS into ECS}
To illustrate how MAS concepts can be used with the ECS pattern, we extended an existing MMOG developed with the ECS architectural pattern. This was our first practical study on leveraging MAS development. The objective of this study was to implement a typical MAS scenario within the system, following the premise that MAS elements can be brought to the chosen architectural pattern. We chose an open-source MMOG called Terasology~\footnote{https://terasology.org/} for our base system. This game is implemented using the ECS architectural pattern, and it is open source, giving us access to the core system implementation. 

Because we used a game platform, we chose a scenario that would use the main concepts involved in an MAS in the context of the game. This meant using the existing structure of entities and components to (i) create our agents, (ii) create a structure that could be used to group these agents into organizations, (iii) create a structure that could allow collective problem-solving, and (iv) making it all without bringing specialized knowledge from the MAS domain.

From the game perspective, the main problem was to establish collective problem-solving. Using an ECS implementation means isolating the logic from both data and entity identification, which means that all group-related functionalities must be handled by system-level logic. In addition, because we wanted to explore concepts related to deliberative MAS, we decided to implement agents using the notion of behaviors. Popular games already use \textit{behavior trees}~\cite{shoulson2011parameterizing} to coordinate characters in-game, allowing this structure to be used without additional MAS-specific concepts and implementation.

With these considerations in mind, we implemented the aforementioned structures. The agents were described in JSON files. Each file described the agent functionalities, which were implemented as Components. These Components included a base specification for all agents, such as rendering attributes (color, skeletal structure, etc.), the organizations (groups) to which the agent belongs, and a reference to a specific behavior tree. An example of an agent descriptor file is presented below:

\begin{verbatim}
{
 "parent" : "baseAgent",
 "Behavior" : {
    "tree" : "stray"
  },
 "Groups" : {
  "groups" : [ "Rebels", "Green" ]
 }
}
\end{verbatim}

In this agent file, the \texttt{parent} description specifies a base for all agents, including their in-game rendering attributes (color, skeletal structure, etc.). The \texttt{Behavior} description specifies a specific behavior tree in the form of a JSON-like file (there is no need to implement a specific behavior tree within the source code). All behaviors were described modularly in pre-defined blocks (similar to a workflow). Finally, the \texttt{Groups} description specifies the organizations (groups) to which the agent belongs (\texttt{Rebels} and \texttt{Green}). Similarly to agents and behavior trees, groups are also described by JSON-like files:

\begin{verbatim}
{
  "groupLabel": "Rebels",
  "syncBehavior": "false",
  "behaviorTree": "flockOrAttack"
}
\end{verbatim}

Besides the identification (\texttt{groupLabel}), the other two parameters refer to the need for synchronized behavior (where all agents act in unison) and the behavior tree that must be imposed on all agents within the group.

Because an ECS Entity does not hold any data other than its own unique identification, its descriptions are in the form of a set of Components. Each Component within the system is implemented using the programming language used by the system (in this case, Java). While logic implementation must be performed within the source code, it is unnecessary to incorporate an agent programming language or declarative expressions of any sort (such as intentions). The most complex notions related to behavior trees (such as how they are interpreted) are encapsulated within a structure that uses a descriptive text file as a reference. Agents can be found solely by their component properties, and communication between them is handled by the system architecture (without needing an ACL). All functionalities can be described as components and the logic behind them. 

This implementation is publicly available as part of Terasology's source code, and was funded as an individual project in the 2019 edition of Google Summer of Code~\footnote{https://summerofcode.withgoogle.com/archive/2019/projects/}.
%%End of 6.1
We adopted the ECS architectural pattern for two reasons. The first one was related to its use and application. Despite their size, MMOGS using this pattern have been deployed in several configurations, from centralized to Web-based DS. Different aspects of more complex DS (e.g., scalability and resource management) can be used without requiring extensions or additions to the pattern. Because it is a data-oriented pattern, every logical requirement (directly related to the agent-related knowledge necessary for each MAS) can be designed and implemented within the pattern. %%%% 
However, existing MASs that follow typical object-oriented patterns have already shown that it is possible to implement any necessary agent capability~\cite{mascardifantastic}, although they do not necessarily adhere to the original architectural pattern. Although using other general-purpose DS patterns could still showcase the same agent capabilities, we decided to choose a fresh approach in this area.

The second reason relates to the "data-oriented" aspect. This characteristic is especially favorable if we want to bring MAS concepts into the Web DS context because most agent-specific techniques and capabilities are based on data structure and processing. Again, while it is not impossible to do the same with other architectural patterns, using ECS makes it easier to encapsulate the agent capabilities within a system.    

If we consider all the possible MAS elements that can be introduced into the system design using this or any DS architectural pattern, they should follow the same rules: they should not break the pattern and should be encapsulated or isolated as much as possible. Otherwise, we would end up reverting to an MAS architecture that uses DS elements, not the opposite. 

As the next step in our study, we decided to use the ECS architectural pattern to design an MAS framework. We showed that using an existing ECS implementation makes it possible to build a community of agents. However, adapting an already existing ECS-based system has several disadvantages. For example, it would be difficult to use the same game system used in our case study to control traffic lights in the real world. In addition, because we would be using the system for a different purpose from gaming, it would contain many undesirable software components (e.g., anything related to 3D rendering).

At this point, we will not detail the design of the framework \textit{per se}, as this is not necessary within the scope of the present paper. However, we used it in the next part of our study, as described in the following paragraphs.

\subsection{Leveraging the Curve}
Bringing Web technologies into the MAS context is an intuitive approach. However, this creates a disadvantage from the overall engineering perspective. While Web technologies are widely available and can be easily learned by developers worldwide, the same does not apply to MAS theory. Any project implementing an MAS requires considerable specialized knowledge from the developers involved, which poses a challenge to the large-scale adoption of MAS~\cite{Mascardi:2019:EMS:3310013.3322175}.

With this in mind, our second practical study in this work was to use the concepts from the first study to teach students without any previous knowledge of agent theory on how to implement a MAS. The idea behind this study was simple: we concluded that it was \textit{possible} to create a framework using an existing DS architectural pattern, but we wanted to determine if it was \textit{necessary}. For this purpose, we divided our efforts into two parts. The first was to use the previously designed framework to allow the students to encapsulate as many concepts as needed for their projects within the same common platform. The second part was to create a different version of the same course without using an ECS-based framework. 

For the first part, we used the existing MAS introductory course offered by USP, Brazil (PCS-5703: multi-agent Systems)~\footnote{https://edisciplinas.usp.br/enrol/index.php?id=83096}. We also used this course as the basis for defining our minimal conceptual set of agent-related concepts, implemented within the framework used by students. At the end of the course, we conducted a guided usability experiment aimed at our implemented framework. This project was submitted and approved by the Ethics Committee~\footnote{https://plataformabrasil.saude.gov.br} under the reference CAAE 40985420.0.0000.5390.

The guided experiment included questionnaires to evaluate the initial proficiency of the students regarding MAS and programming. One of the prerequisites for this course was previous exposure to the Java programming language since the students would have to use a Java-based implementation of the framework we previously designed. Using these questionnaires, we were able to profile the students taking the course as follows:

\begin{itemize}
    \item[$\bullet$] Software development proficiency: low (13\%), medium (56\%), and high (31\%). 
    \item[$\bullet$] Professional software development experience: none (13\%), 6 months or less (6\%), between 6 months and 2 years (31\%), and 2 years or more (50\%).
    \item[$\bullet$] JAVA development proficiency: none (19\%), low (44\%), medium (31\%), and high (6\%).
    \item[$\bullet$] Development experience using APIs: Yes (37\%), No (63\%).
    \item[$\bullet$] Development experience using Distributed Systems: Yes (19\%), No (81\%).
\end{itemize}

During the course, we required students to implement an MAS for an auction mechanism. Our DS-based implementation allowed them to work with all topics covered by our proposed minimal conceptual set. The required auction mechanism was based on a tutorial scenario for an existing MAS framework, JaCaMo~\footnote{https://jacamo-lang.github.io/}. The scenario in question is called "house building"~\footnote{https://sourceforge.net/p/jacamo/code/HEAD/tree/tut/house-building/}. We chose it for the required implementation because it is already a proven introductory scenario for building MAS, and it encompasses all theoretical topics within our minimal conceptual set.

In the second part of our study, we decided to create a different course for the students using an engineering-based approach. The new course was offered in 2023 at USP, Brazil, under the name "PCS-5045: MAS Engineering I"~\footnote{https://edisciplinas.usp.br/course/view.php?id=110364}. We have amended the previous Ethics Committee submission (under the same reference), which was approved before we started our lectures. 

For this course, the requisites were pretty much the same, except for one: we did not ask for any Java proficiency since the students would not use any specific framework. We also asked them to design an MAS to solve a particular problem, chosen from a list of possible cases based on real-world scenarios. While the scenario used in the previous course was ideal for an MAS implementation, it was also \textit{leading} in terms of system requirements - in the sense that it was conceived as part of an implementation tutorial. For this reason, we decided to present the students with a set of scenarios that \textit{could} be solved using MAS.

The students taking the new course were profiled as follows:

\begin{itemize}
    \item[$\bullet$] Software development proficiency: low (23\%), medium (33\%), and high (44\%).
    \item[$\bullet$] Professional software development experience: none (11\%), 6 months or less (11\%), between 6 months and 2 years (33\%), 2 years or more (45\%).     
    \item[$\bullet$] Development experience using APIs: Yes (56\%), No (44\%).
    \item[$\bullet$] Development experience using Distributed Systems: Yes (33\%), No (67\%).
\end{itemize}  

We also used the same minimum conceptual set in this course's lectures. However, since we were not providing a framework for them to use, we presented the basics of the DS and MAS architectural patterns. We also allowed them to use SPADE~\footnote{https://spade-mas.readthedocs.io/en/latest/readme.html} for their final projects. SPADE is a minimal MAS framework written on top of Python that provides the basic structures for a MAS. However, because we intended to detach their development experience from specific MAS-related tools as much as possible, we only allowed them to use the basic agent structure provided by the framework. This decision was taken considering (i) the duration of the course, (ii) the complexity of the proposed projects, and (iii) the fact that the basic concept of an "agent" - as implemented within the SPADE framework - is already used in the DS area, and it is implemented by widely used DS-specific frameworks such as Akka~\footnote{https://akka.io/} and Orleans~\footnote{https://learn.microsoft.com/en-us/dotnet/orleans/}.

\subsection{Learning Outcomes}

The expected learning outcomes from both courses were:

\begin{itemize}
    \item[$\bullet$] Knowledge: Understanding the key characteristics of agents and multi-agent systems.
    \item[$\bullet$] Skills: The ability to engineer basic multi-agent systems to solve cooperative problem-solving scenarios.
\end{itemize}  

To measure these outcomes, students were evaluated according to the following method:

\begin{itemize}
    \item[$\bullet$] Reading: Each week, students received an article related to the topic of the upcoming class. They were required to read the article and submit a summary of its content the following week.
    \item[$\bullet$] Practical work: During the course, the students were required to complete three to four hands-on projects by applying the concepts studied in class. These projects were cumulative, with each new project building on the previous project.     
    \item[$\bullet$] Seminar: At the end of the course, the students were required to present a seminar on the project they developed, explaining how the concepts studied in class were applied.
\end{itemize}  

\begin{figure}[htbp]
\centering
\includegraphics[width=0.5\textwidth]{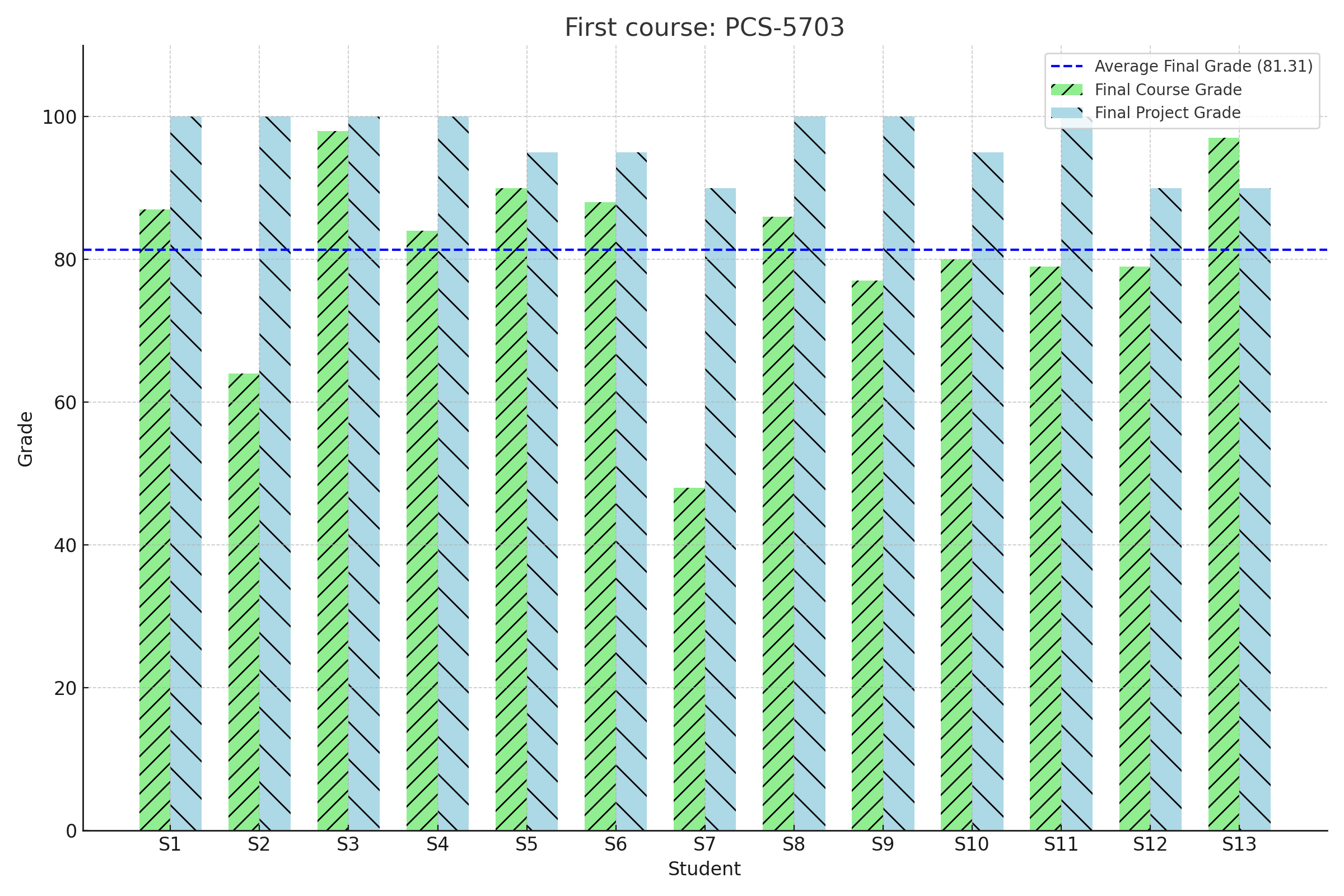}
\caption{Learning outcome achieved: Grades for the course PCS-5703 (Source: author)}
\label{fig:grades_pcs5703}
\end{figure}

\begin{figure}[htbp]
\centering
\includegraphics[width=0.5\textwidth]{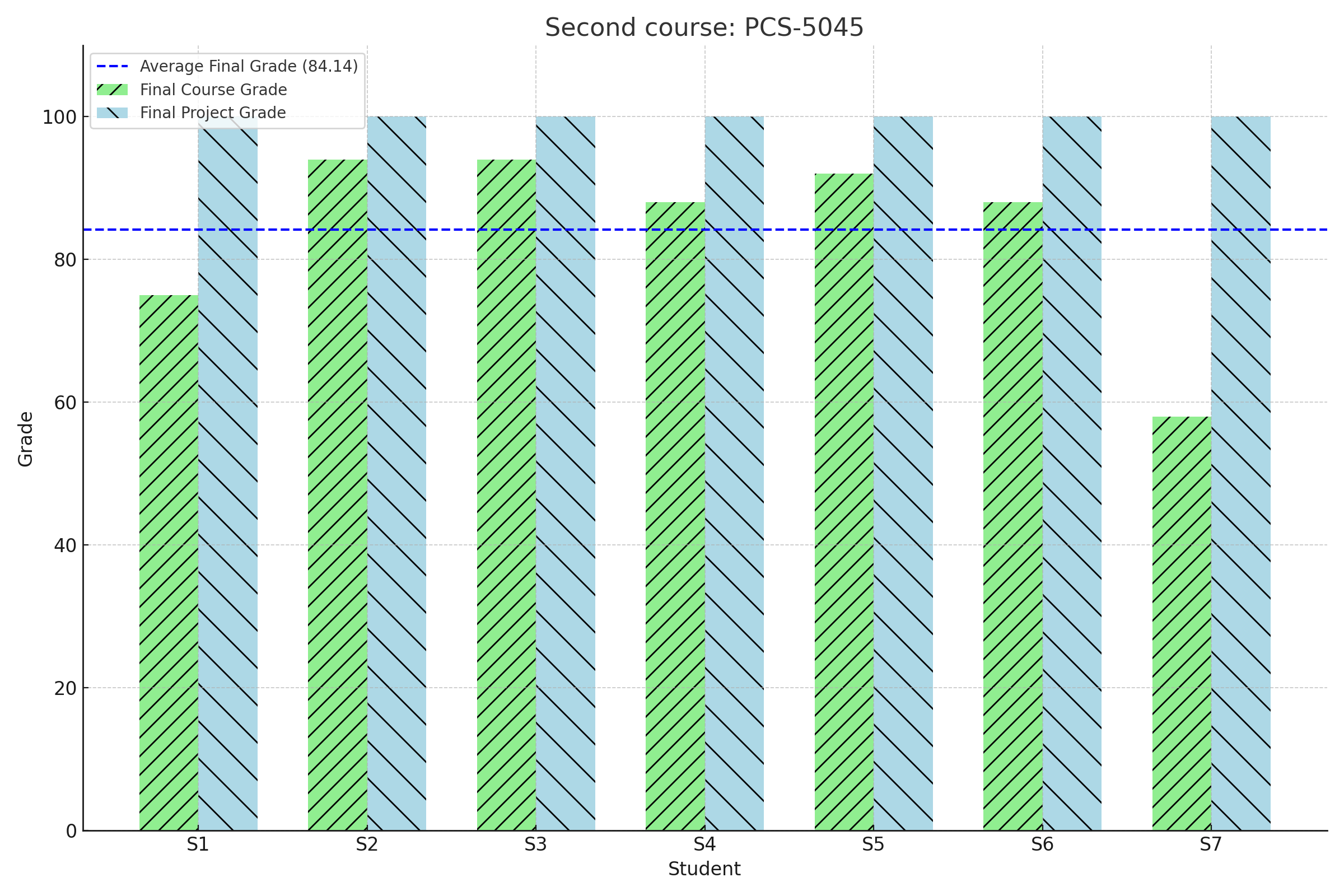}
\caption{Learning outcome achieved: Grades for the course PCS-5045 (Source: author)}
\label{fig:grades_pcs5045}
\end{figure}

Figures~\ref{fig:grades_pcs5703} and~\ref{fig:grades_pcs5045} show the final grades obtained by students in both courses. In each graph, two different grades (Y-axis) are shown for each student (X-axis). The left column shows the final grade obtained by the students in the course, and the left column shows the grade obtained by the students in the final project.

The average final grade for the first course was 81.31; for the second course, it was 84.10 (out of 100). It is interesting to note that the grade obtained by each student for the final project was always higher than the average final grade. For the first course, none of the students scored lower than 90 in the final project, whereas in the second course, all students scored 100 out of 100 in the final project.

These grades can be analyzed from the perspective of each course. In the first course, the students were taught using our designed framework. They also used this same framework in their final project. In the second course, the students were oriented towards DS tools and techniques. If we consider that in both courses, (i) the agent-related topics evaluated in each project were the same and (ii) the programming proficiency of the students was, on average, the same, we can infer that the newly adjusted course had a positive impact on their learning outcomes. 

In terms of the benchmark, we used the original course (PCS-5703), as it was taught in 2016. In this instance of the course, while the syllabus and topics were the same as we used, the students did not have access to the framework we developed. Consequently, the practical project was based on the MAS-specific framework JaCaMo: 

\begin{figure}[htbp]
\centering
\includegraphics[width=0.5\textwidth]{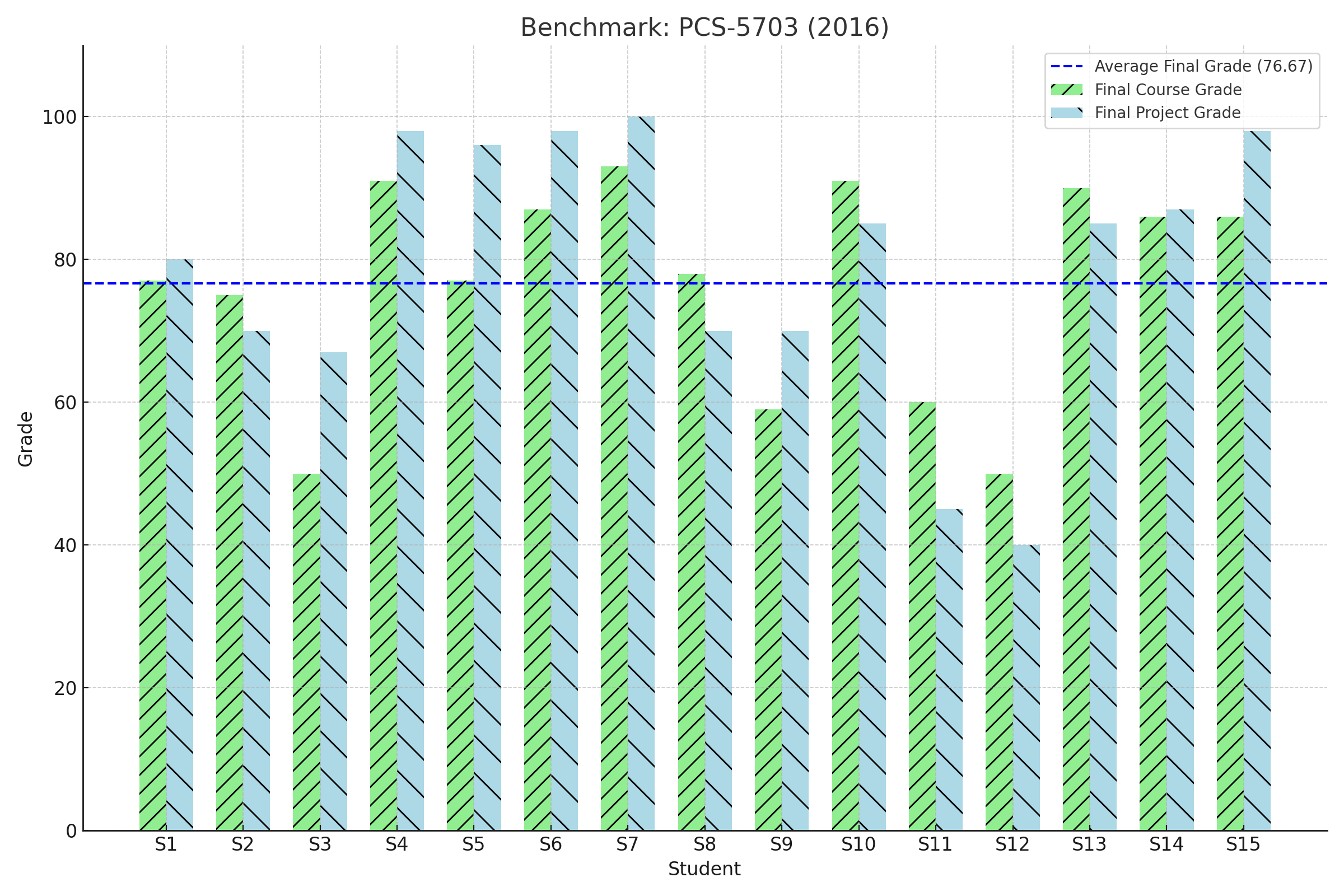}
\caption{Benchmark: Past grades for the course PCS-5703 (Source: author)}
\label{fig:grades_pcs5703_2016}
\end{figure}

When we used the grades obtained by the students in a past instance of the course PCS-5703 (Figure~\ref{fig:grades_pcs5703_2016}) as a benchmark, the immediate observation was that the framework we developed and used in the latest instance of the same course also had a positive impact on the learning outcome.

However, there are a few limitations to the analysis of these courses. The first refers to the benchmark grades: although the course topics were the same, the scope of the practical project conducted by the students was different. This is partly due to the steep learning curve of using MAS-specific frameworks in general. While we understand that part of the merit of using a custom-made framework is to avoid this learning curve, a proper benchmark should involve the same scope for practical projects in both instances of the course. 

The second limitation concerns the previous exposition of students on the course topic. In both courses, one or two students had contact with agent-specific frameworks after the course as part of their master's or doctorate topics. Individual feedback from these students regarding using different tools (custom framework or DS tooling) was positive. However, we would need more instances of the same courses and control groups to properly measure the impact of the changes we introduced.

Finally, we collected feedback from students in both courses at the end of the term (during their seminar presentation). This feedback was unanimously positive: the students felt it was much easier to understand agent theory by applying the concepts learned to a practical project during the course.

\section{Discussion and Future Work}
\label{sec:Discussion}
In the present work, we trace a historical parallel between the evolution of Distributed Systems and Multi-Agent Systems research fields, pointing out the common origins and similarities between them. We compared architectural styles and patterns used in the implementation of MAS and DSs, with an emphasis on the communication mechanisms typically used. 

This parallel was then used to point out that while web developers can seamlessly implement distributed systems, the same is not valid for multi-agent systems. Implementing an MAS requires specific knowledge of agent-related theory, even when using available agent-oriented frameworks and libraries. This was pointed out in the past as one of the possible reasons impeding MAS from being widely adopted by the industry~\cite{Mascardi:2019:EMS:3310013.3322175}. Therefore, we decided to follow the example of the Web and introduce concepts from MAS theory into the DS context. 

After establishing an architectural parallel between the MASs and DSs, we created a minimal set of MAS-related concepts that could be introduced into the DS domain. Once this set was created, we tested these concepts within the context of our proposal. For this, we took several cumulative steps. First, we chose a DS architectural pattern representative enough for our study, considering the architectural comparison previously made and the relevant complexity considerations. The pattern chosen was then used in two experiments. The first involved implementing agent-related concepts from our minimal set into an existing system. This experiment was performed and validated as an accepted result in the 2019 edition of the Google Summer of Code.

The second step was to use the chosen pattern to design a framework that accommodated all concepts contained in our minimal set. Designing such a framework was important for studying the architectural and technical limitations of our approach. This design (and its subsequent implementation) will be discussed in detail in a separate publication.

The next step was to validate the conceptual set more broadly. For this purpose, we have taught two graduate courses at the university. With a more conceptual approach, the first used the previously designed framework and required students with no previous MAS-related knowledge to implement a specific MAS. The second had a practical approach, and the students (with no prior knowledge of MAS-related theory) were required to design and implement an MAS to solve a specific problem. In both cases, the students were only provided with agent theory contained within our proposed minimal conceptual set. 

We profiled the students taking both courses and found that their experience and proficiency in software development were similar in both cases. However, students taking the second course had more previous contact with APIs and Distributed Systems. 

In both courses, all students were able to implement the required MASs within the DS domain. In the first course, we used a scenario adapted from an implementation tutorial for JaCaMo. Because we were using our framework implementation, we chose this scenario as part of a proven implementation tutorial within the MAS domain. In the second course, however, we asked the students to choose between multiple real-world scenarios with problems that could be solved in several ways, including implementing MAS. We wanted to understand how much of the recently learned agent theory would be used by the students in their implementation.

The learning outcomes of both the courses were positive. In addition to the high-grade average observed in both cases, the final feedback from the students from both courses at the end of the term was unanimously positive: they felt it was much easier to understand agent theory by applying the concepts learned in a practical project along the course. We intend to offer these courses again in the next few years.

In light of this study and the nature of all implementations, we concluded that our approach was successful, although limited by the number of experiments. Considering the nature of our approach, we can use it not only to develop new MASs but also to bring MAS-specific capabilities into existing systems. Establishing a minimum MAS conceptual set and using it in conjunction with existing distributed technologies allows for the development of specialized MAS without depending on specific MAS frameworks. At the same time, it will enable software engineers to incorporate the latest AI advancements and techniques into their work without losing touch with existing agent research and without abstaining from using DS-specific frameworks or techniques.

In terms of broader implications, we were also able to observe a few interesting findings in this study. The first relates to the existing synergy between MAS and DS concepts. By analyzing the core elements taught in agent-related courses, it was clear that most of these concepts were common to DS areas. Although this can be easily explained by the common origin of both research fields, it does not alter the fact that there is a fertile space for cross-pollination between DS and MAS. By using DS-specific tools and techniques, MAS engineers can benefit from topics already well-discussed and treated in the DS field, such as fault tolerance, load balancing, and scaling. Simultaneously, problems usually belonging to the MAS domain can be solved by applying DS techniques. The concept of coordination in MAS is well-aligned with the peer-to-peer (P2P) architecture~\cite{aberer2005essence} in DS, and agent modeling and communication can benefit from the use of microservices~\cite{thones2015microservices}. The scope of the present work does not include a full study of which of these techniques has already been incorporated across both research fields. However, considering the present state of both fields, we believe that introducing MAS problems into the DS domain would yield better practical results than doing the opposite. We intend to explore this perspective in future studies.

It is also important to note that this approach can be generalized to other AI and engineering systems beyond the MAS and DS domains. Some of these are already being explored, such as integrating computer vision and autonomous systems seen in self-driving cars or combining deep learning and edge computing in smart cities. In a broader scenario, MAS and DS can be used in conjunction with other AI and engineering techniques to produce more sophisticated systems. The key and arguably most important benefit of cross-pollination between AI and engineering fields is leveraging the problems already solved in the engineering field with the latest AI advancements. This perspective is part of what fuels our research, and we believe many opportunities have yet to be fully explored.

\appendices
\section{\break Comparative Analysis of Syllabi in multi-agent Systems Courses}
\label{app:books}

This appendix contains a detailed analysis of the usage of four major textbooks in multi-agent systems (MAS) courses, as referenced in their syllabi. We analyzed 26 graduate courses taught in universities worldwide, comparing the usage of each book in Brazilian universities, international universities, and globally (altogether). 

\subsection{Textbooks for MAS Grad Courses}
By analyzing the course syllabi, we found that the prevalence of the following textbooks (editions and publication years may vary):
\begin{enumerate}
    \item \textbf{Multi-agent Systems: Algorithmic, Game-theoretic, and Logical Foundations}, by Yoav Shoham and Kevin Leyton-Brown.
    \item "\textbf{An Introduction to multi-agent Systems}," by Michael Wooldridge.
    \item "\textbf{Artificial Intelligence: A Modern Approach}," by Stuart Russell and Peter Norvig.
    \item "\textbf{Multi-agent systems: A modern approach to distributed artificial intelligence}," by Gerard Weiss.
\end{enumerate}

\subsection{Data Sources}
The following universities and courses were surveyed in the analysis:

\subsubsection{Brazilian Universities}
\begin{itemize}
    \item \textbf{University of São Paulo (USP)}: Course on \textit{Multi-Agent Systems}.
    \item \textbf{Federal University of Pernambuco (UFPE)}: Course on \textit{Intelligent Agents}. 
    \item \textbf{Federal University of Santa Catarina (UFSC)}: Course on \textit{Multi-Agent Systems}.     
    \item \textbf{University of Brasília (UnB)}: Course on \textit{Multi-Agent Systems}. 
    \item \textbf{Federal University of Bahia (UFBA)}: Course on \textit{Autonomous Agents and Multi-Agent Systems}. 
    \item \textbf{Federal University of Maranhão (UFMA)}: Course on \textit{Multi-Agent Systems}. 
    \item \textbf{Federal University of Rio Grande do Sul (UFRGS)}: Course on \textit{Autonomous Agents and Multi-Agent Systems}. 
    
\end{itemize}

\subsubsection{International Universities}
\begin{itemize}
    \item \textbf{University of Western Australia}: Course on \textit{Intelligent Agents}.
    \item \textbf{KU Leuven}: Course on \textit{Multi-Agent Systems}.
    \item \textbf{Technical University of Denmark}: Course on \textit{Artificial Intelligence and Multi-Agent Systems}.
    \item \textbf{University of Bologna}: Course on \textit{Multi-Agent Systems}.
    \item \textbf{Maastricht University}: Course on \textit{Agents and Multi-Agent Systems}.
    \item \textbf{University of Amsterdam}: Course on \textit{Multi-Agent Systems}.
    \item \textbf{Norwegian University of Science and Technology (NTNU)}: Course on \textit{Multi-Agent Systems and Game Theory}.
    \item \textbf{University of Oslo}: Course on \textit{Multi-Agent Systems}.
    \item \textbf{University of Auckland}: Course on \textit{Intelligent Software Agents}.
    \item \textbf{University of Algarve}: Course on \textit{Multi-Agent Systems}.
    \item \textbf{University of Lisboa}: Course on \textit{Multi-Agent Systems}.
    \item \textbf{University of Minho}: Course on \textit{Agents and Multi-Agent Systems}.
    \item \textbf{University of Porto}: Course on \textit{Multi-Agent Systems}.
    \item \textbf{Barcelona School of Informatics (FIB)}: Course on \textit{Introduction to Multi-Agent Systems}.
    \item \textbf{Swiss Federal Institute of Technology Lausanne (EPFL)}: Course on \textit{Intelligent Agents}.
    \item \textbf{University of Liverpool}: Course on \textit{Multi-Agent Systems}.
    \item \textbf{University of Southampton}: Course on \textit{Intelligent Agents}.
    \item \textbf{Carnegie Mellon University}: Course on \textit{Multi-Agent Systems}.
    \item \textbf{Harvard University}: Course on \textit{Multi-Agent Systems}.
\end{itemize}

\subsection{Comparative Usage of MAS Books in Courses}
The following table summarizes the usage of the three main textbooks in Brazilian and international universities, along with their global usage (a combination of both Brazilian and international data).

\begin{table}[htbp]
\caption{Comparative Usage of the prevalent MAS Books by region: Local (Brazil), International (Intl.), Global}
\label{tab:comparative_region}
\centering
\begin{tabular}{|l|c|c|c|}
\hline
\textbf{Book} & \textbf{Brazil (\%)} & \textbf{Intl. (\%)} & \textbf{Global (\%)} \\
\hline
Wooldridge & 85.71\% & 78.95\% & 80.77\% \\
Shoham \& Leyton-Brown & 14.29\% & 36.84\% & 30.77\% \\
Russell \& Norvig & 42.86\% & 57.89\% & 53.85\% \\
Weiss & 71.43\% & 15.79\% & 30.77\% \\
\hline
\end{tabular}
\end{table}

\subsection{Analysis and Graphics}
Our analysis shows that \textit{Wooldridge} is used in the majority of courses related to multi-agent systems. Although \textit{Russell and Norvig}'s book is a common reference for AI courses, it does not appear in all syllabi. Also, the books by \textit{Shoham and Leyton-Brown} and \textit{Weiss} show different levels of usage in different regions. 

\begin{figure}[htbp]
\centering
\includegraphics[width=0.5\textwidth]{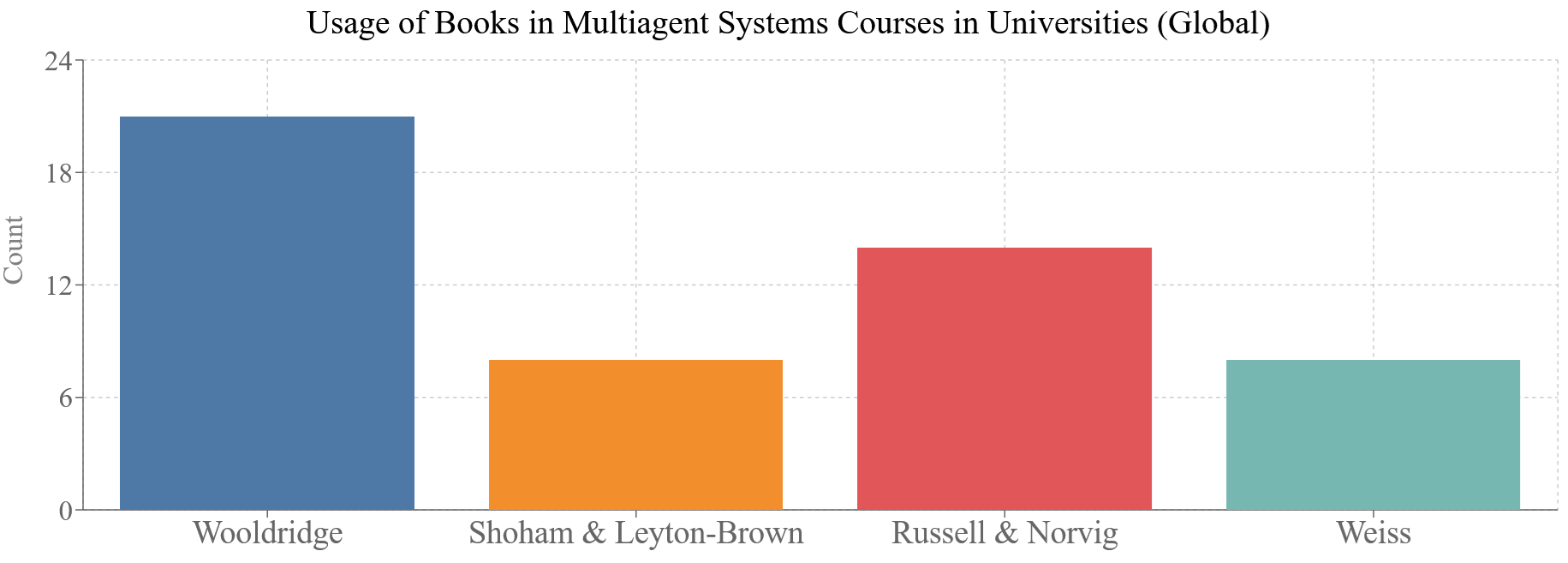}
\caption{Usage of Books in Multi-Agent System Courses Globally (Source: authors)}
\label{fig:global_comparison}
\end{figure}

For a graphical representation, please refer to Figure \ref{fig:global_comparison}.

\bibliographystyle{unsrt}
\bibliography{references}  

\begin{IEEEbiography}[{\includegraphics[width=1in,height=1.25in,clip,keepaspectratio]{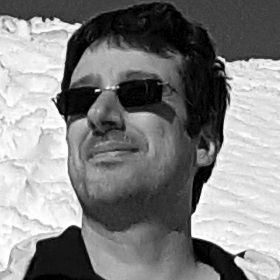}}]{Arthur Casals} received the B.S. degree in Computer Engineering from ITA - Instituto Tecnológico de Aeronáutica, São José dos Campos, SP, Brazil, in 2004 and the Ph.D. degree in Informatique, Télécommunications et Électronique from Sorbonne Université, Paris, France, in 2022. He is currently pursuing a Ph.D. degree in Computer Engineering at Escola Politécnica da USP, São Paulo, Brazil. From 2005 to 2018, he assumed different roles in the industry, all related to technology development and its applications. His research interest includes multi-agent systems, distributed systems, and software engineering. 
\end{IEEEbiography}

\begin{IEEEbiography}[{\includegraphics[width=1in,height=1.25in,clip,keepaspectratio]{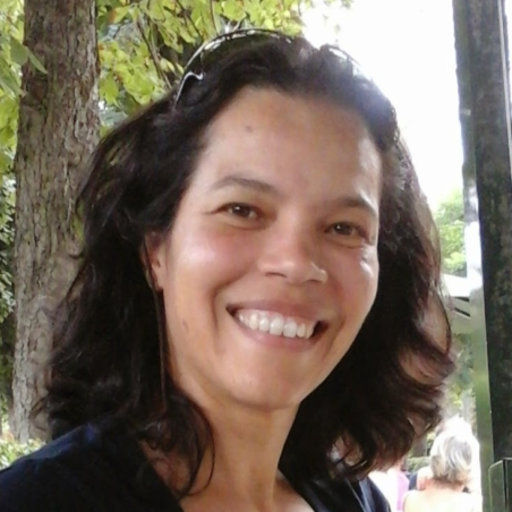}}]{Anarosa A. F. Brandão} holds a Bachelor's degree (1990) and a Master's degree (1994) in Mathematics from the University of São Paulo, and obtained his Ph.D. in Computer Science from the Pontifical Catholic University of Rio de Janeiro (2005). He earned his Livre Docente title from the University of São Paulo in 2021. Her research interests lie in Computer Science, with a focus on Artificial Intelligence and Informatics in Education. Her work primarily centers around agent and multi-agent systems, and accessible web-based learning systems. Since December 2008, she has served as a professor in the Department of Computer Engineering and Digital Systems at the Escola Politécnica of the Universidade de São Paulo. At the same year she became the mother of Leonardo. She was the editor-in-chief of the Brazilian Journal of Informatics in Education (RBIE), published by CEIE-SBC, from 2022 to 2024. Currently, she serves as the President of the Inclusion and Belonging Committee at the Escola Politécnica (2024-2026). From October 2022 to February 2024, she held the position of Coordinator of the Computer Engineering Concentration Area within the Graduate Program in Electrical Engineering at the Escola Politécnica of USP. Additionally, she has been representing the Escola Politécnica at the Brazilian Computing Society (SBC) since March 2021. She also contributes as a member of the advisory committee for the WESAAC series of workshops - Workshop on School of Agent Systems, their Environments, and Applications.
\end{IEEEbiography}

\EOD

\end{document}